\documentclass[11pt]{article}
\usepackage{moriond,epsfig}
\usepackage{url}

\newcommand{\squishlist}{
 \begin{list}{$\bullet$}
  { \setlength{\itemsep}{1pt}
     \setlength{\parsep}{3pt}
     \setlength{\topsep}{3pt}
     \setlength{\partopsep}{1pt}
     \setlength{\leftmargin}{2em}
     \setlength{\labelwidth}{1em}
     \setlength{\labelsep}{0.5em} } }
\newcommand{\squishend}{
  \end{list}  }

\def\be{\begin{equation}}
\def\ee{\end{equation}}
\def\bea{\begin{eqnarray}}
\def\eea{\end{eqnarray}}

\begin{document}
\vspace*{4cm}
\title{Results of charged pions cross-section in proton
  carbon interaction at 31 GeV/c measured with the NA61/SHINE detector.}

\author{Sebastien Murphy \\on behalf of the NA61/SHINE collaboration \url{http://na61.web.cern.ch/}}

\address{University of Geneva, DPNC, 24 Quai E. Ansermet, 1201 Geneva, Switzerland}

\maketitle\abstracts{ Among other goals, the NA61/SHINE (SHINE
  $\equiv$ SPS Heavy Ion and Neutrino Experiment) detector at CERN SPS
  aims at precision hadro-production measurements to characterise the
  neutrino beam of the T2K experiment at J-PARC.  These measurements
  are performed using a 31 GeV/c proton beam produced at the SPS with a
  thin carbon target and a full T2K replica target. Spectra of
  $\pi^{-}$ and $\pi^{+}$ inclusive inelastic cross section were
  obtained from pilot data collected in 2007 \cite{pion_paper} with a
  2 cm thick target (4\% of the interaction length). The SHINE
  detector and its particle identification capabilities are described and
  the analysis techniques are briefly discussed.}

\section{Physics motivation}
In T2K, neutrinos are produced by a high intensity proton beam of 31
GeV/c impinging on a carbon target and producing mesons ($\pi$ and
$K$) from the decay of which the neutrinos are produced. There exist
so far no measurements of hadron inclusive spectra from p+C at 31
GeV/c.  Thus the NA61/SHINE experiment provides a precise measurement
of meson yield production in carbon at the proton beam energy of
interest for T2K. These measurements are used for the T2K neutrino
beam simulation and consequently reduce the systematic uncertainties
of the neutrino energy distribution at the needed level for the
physics goals of T2K \cite{nico}.

\section{The SHINE detector and combined particle identification}
\begin{figure}[h!]
	\centering
        \includegraphics[width=.7\textwidth,height=.25\textheight]{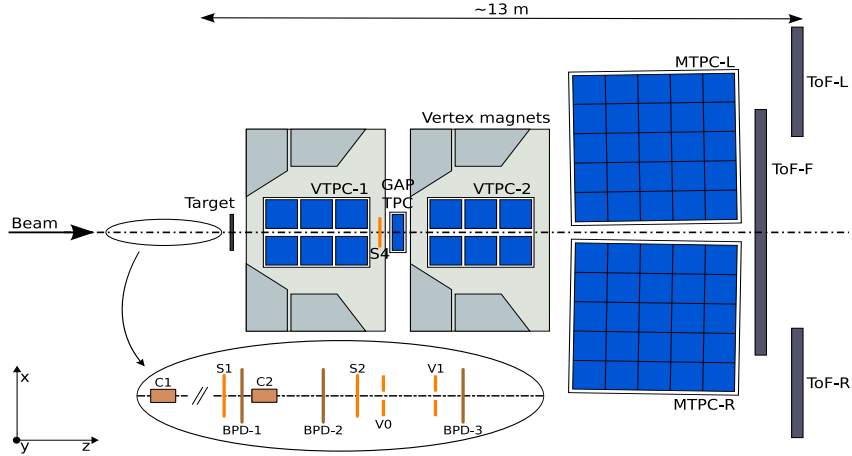}
	\caption{\label{fig:layout}The layout of the NA61/SHINE set-up
          in the 2007 data taking.}
\end{figure}

The set-up of the NA61/SHINE is shown in Fig.~\ref{fig:layout}. The
main components of the NA61 detector were constructed and used by the
NA49 experiment \cite{NA49}. The tracking apparatus consists in four
large volume Time Projection Chambers (TPCs). Two of them, the vertex
TPCs (VTPC-1 and VTPC-2), are located in the magnetic field of two
super-conducting dipole magnets and, two TPCs (MTPC-L and MTCP-R) are
positioned downstream of the magnets, symmetrically on the left and
right of the beam line. The TPCs provide a measurement of charged
particle momenta $p$ with a high resolution. For the 2007 run a new
forward time of flight detector (ToF-F) was constructed in order to
extend the acceptance of the NA61/SHINE set-up for pion and kaon
identification as required for the T2K measurements
\cite{sta_rep_2008}. The ToF-F detector consists of 64 scintillator
bars, vertically orientated, and read out on both sides with Hamamatsu
R1828 photo-multipliers. The resolution of the ToF-F wall is $<$ 120
ps \cite{sta_rep_2008} which provides a 5 $\sigma$ $\pi$/K separation
at 3 GeV/c. It is installed downstream of the MTPC-L and MTPC-R,
closing the gap between the ToF-R and ToF-L walls. The ToF-F provides
full acceptance coverage of the T2K phase-space (parent particles
generating a neutrino which hit the far detector).

\begin{figure}
\begin{center}
\includegraphics [width=0.31\linewidth]{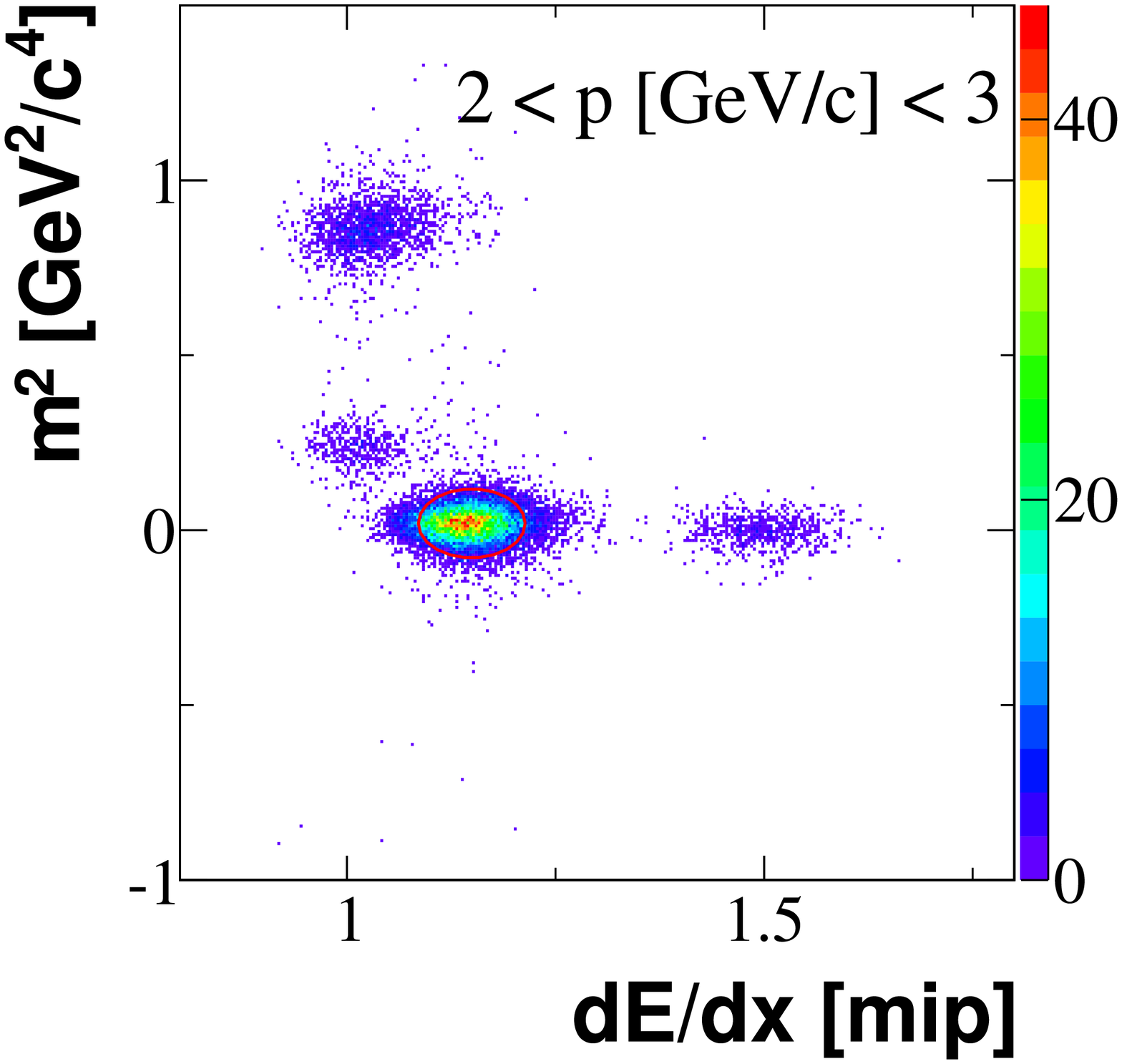}
\includegraphics [width=0.31\linewidth]{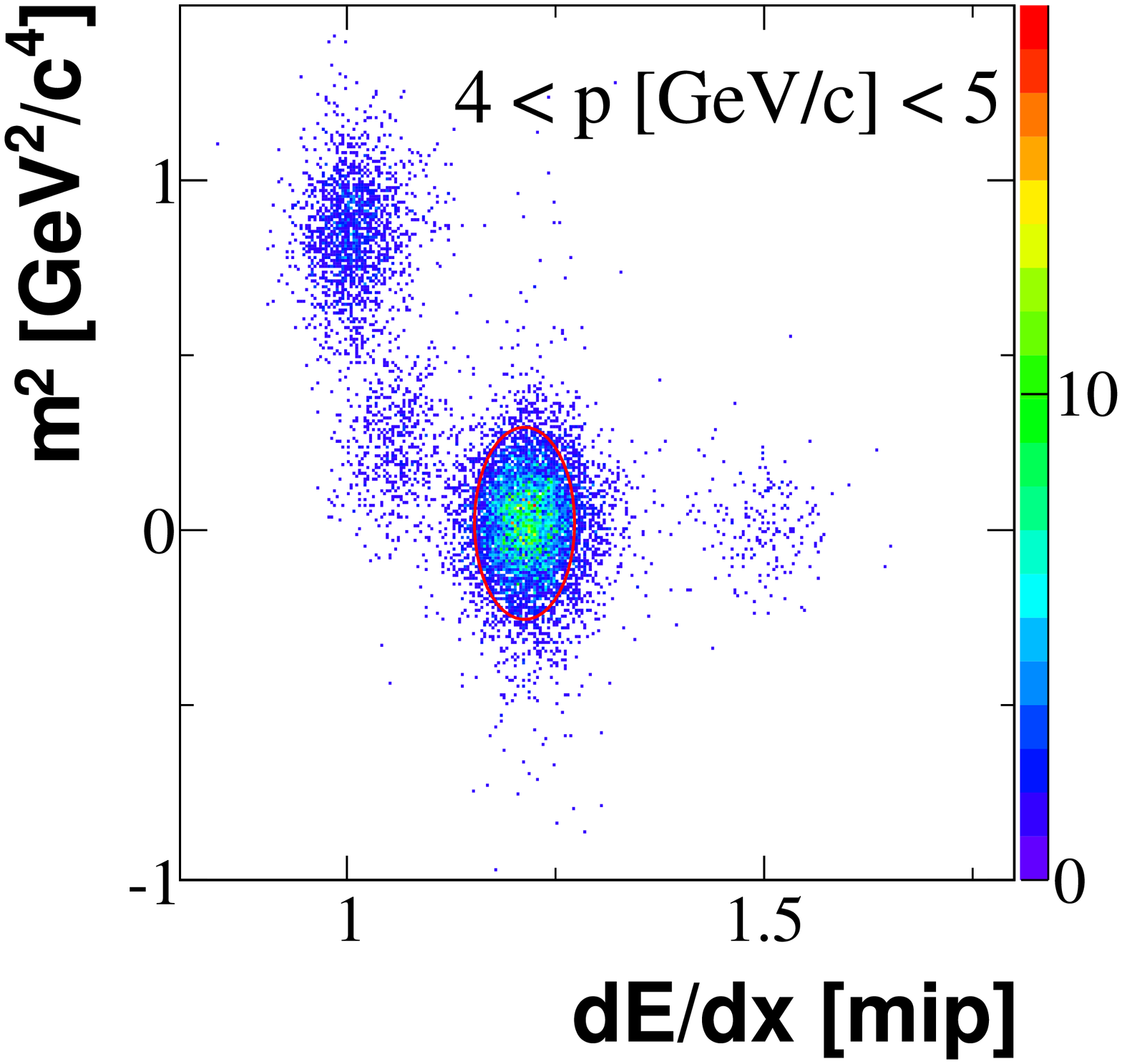}
\includegraphics [width=0.31\linewidth]{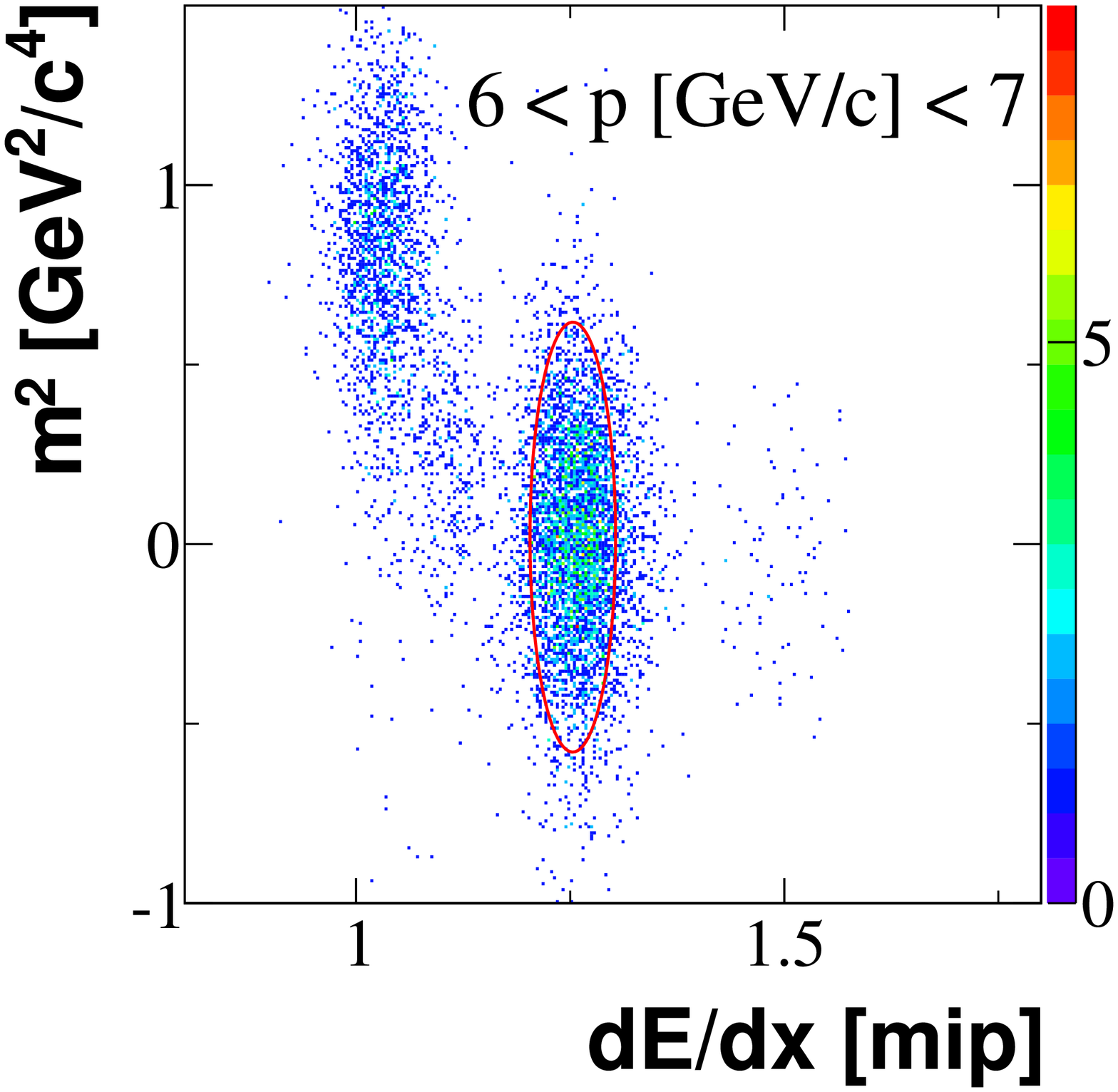}
\caption{
  Examples of two-dimensional $m^2$--$dE/dx$ plots for
  positively charged particles in three
  momentum intervals indicated in the panels. 
  $2\sigma$ contours around fitted pion peaks are shown.
  The left and middle plots correspond to the $dE/dx$ cross-over
  region while the right plot is at such a high momentum that the ToF-F
  resolution becomes a limiting factor. The combination of both
  measurements provides close to 100\% purity in the pion
  selection over the whole momentum range.
}
\label{fig:tof-dedx}
\end{center}
\end{figure}

As demonstrated in Fig.~\ref{fig:tof-dedx}, high purity particle
identification can be performed by combining the $tof$ and $dE/dx$
information over the whole momentum range needed for T2K. Moreover, in
the momentum range 1--4~GeV/c, where $dE/dx$ bands for different
particle species overlap, particle identification is in general only
possible using the $tof$ method. In each ($p$, $\theta$) bin the
bin-by-bin maximum likelihood method was applied to fit yields of
$\pi^+$ and $\pi^-$ mesons. The pion yields were calculated summing
all particles within 2$\sigma$ around the fitted pion peak.

\section{Charged pion cross sections}

The differential inclusive inelastic cross section
$\frac{d\sigma_{inel}}{dp}$ are extracted
using three independent analysis: \squishlist
\item $\pi^{+}$ and $\pi^{-}$ spectra identified with $dE/dx$ below 800 MeV/c
  \cite{magda_proc}.
\item $\pi^{-}$ spectra from a so called h-minus analysis in which all negative
  tracks were selected and yields were extracted from a global Monte
  Carlo factor \cite{tomek_proc}. 
\item $\pi^{+}$ and $\pi^{-}$ yields identified with the combined
  $tof-dE/dx$  method \cite{Sebastien}.
\squishend

\begin{figure}[h!]
	\centering
        \includegraphics[height=.19\textheight,
        width=.43\textwidth]{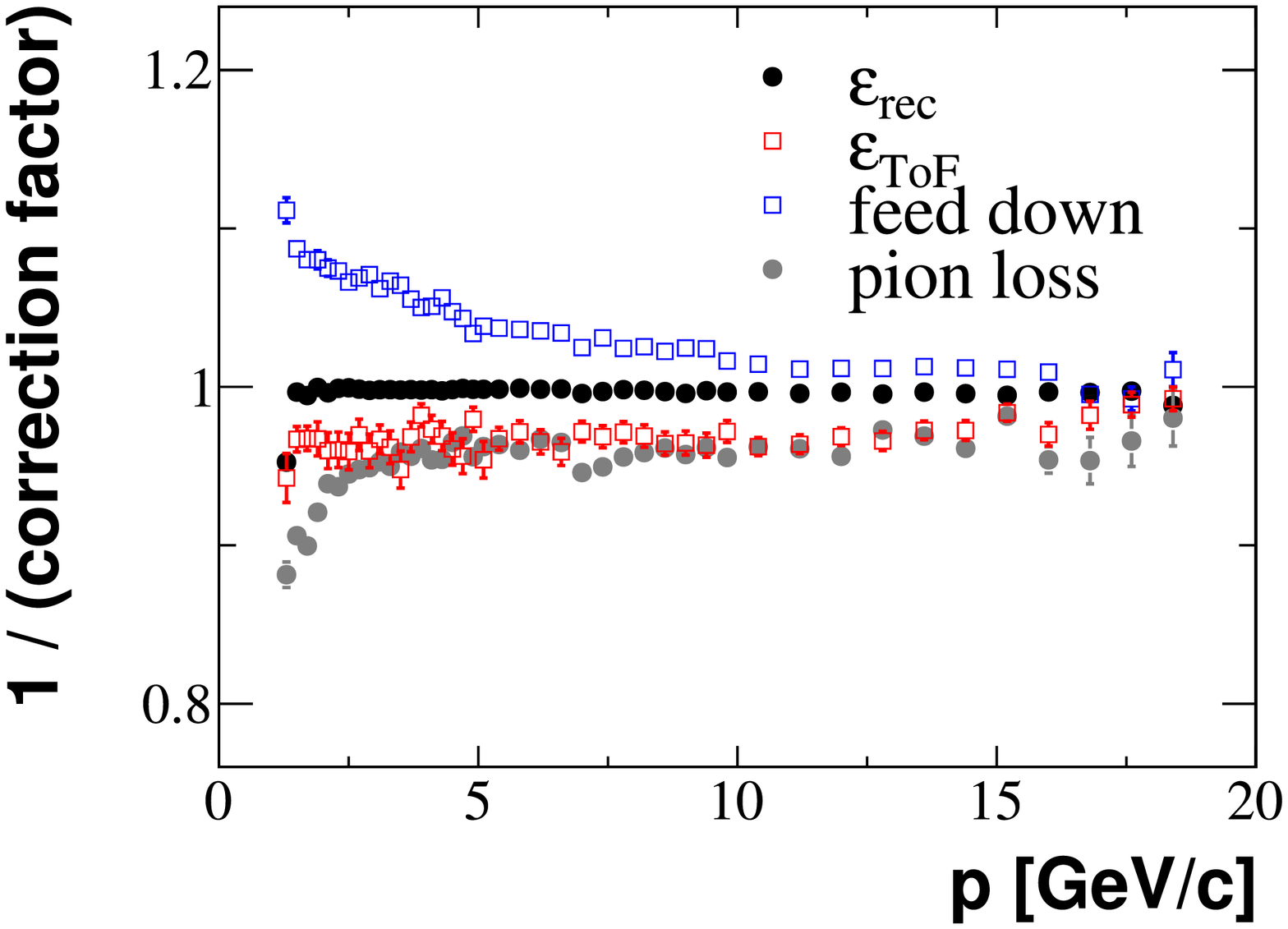}
        \includegraphics[height=.19\textheight,
        width=.43\textwidth]{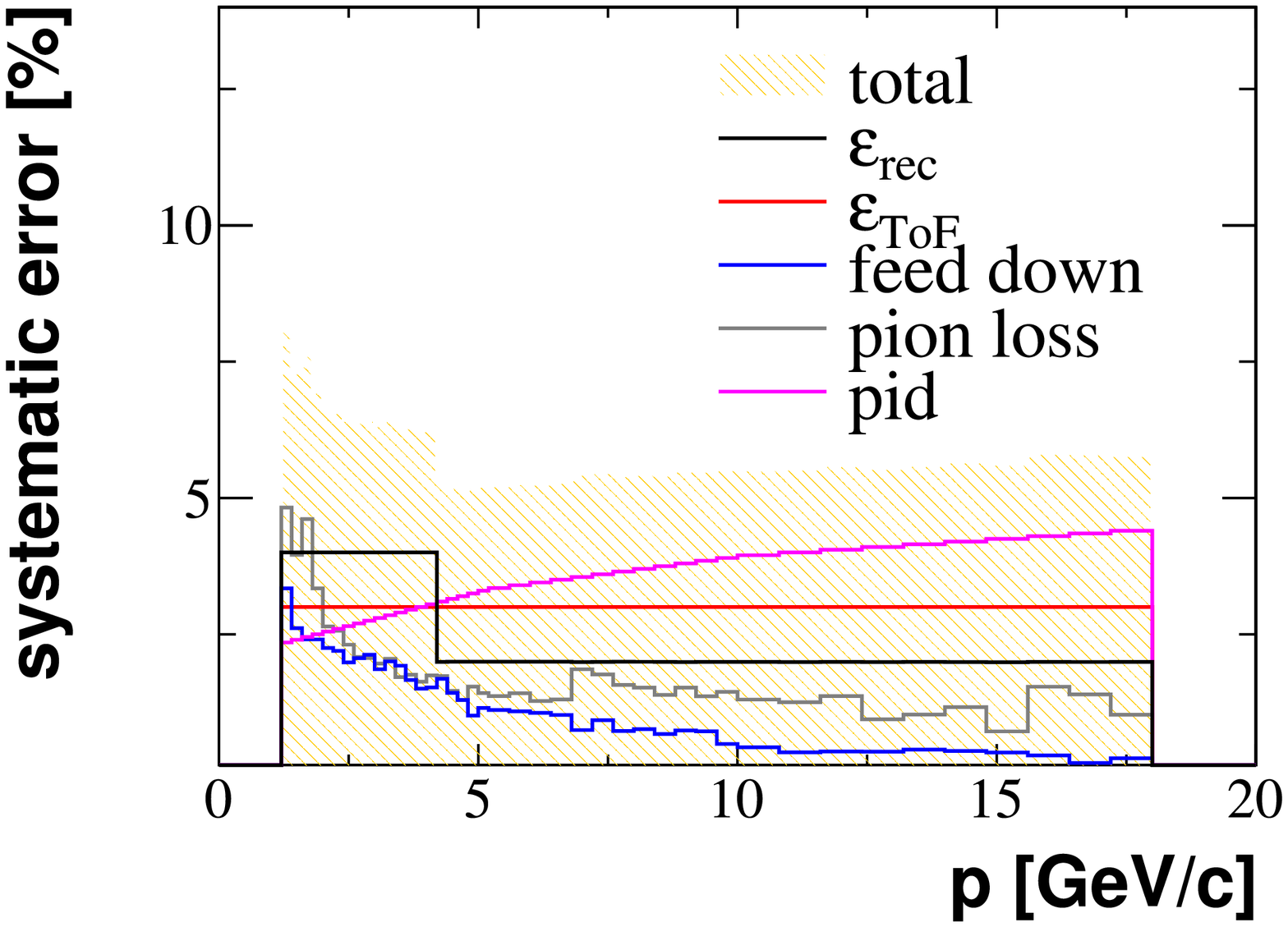}
	\caption{\label{fig:tof-dedx-cor} Example of momentum
          dependence of the inverse correction factor (left) and
          systematic errors (right) for the $tof-dE/dx$ analysis for
          positively charged pions in the polar angle interval [40,60]
          mrad. $\epsilon_{rec}$ and $\epsilon_{tof}$ are the
          efficiencies of the reconstruction and of the ToF-F,
          respectively. The feed-down correction accounts for pions
          from weak decays which are reconstructed as primary
          particles, while the pion loss accounts for pions lost due
          to decays or secondary interactions.}
\end{figure}

All pion yields were corrected with the help of the NA61 Geant3 based
Monte-Carlo. The following effects have been accounted for:
geometrical acceptance of the detector; efficiency of the
reconstruction chain; decays and secondary interactions; ToF detection
efficiency; pions coming from Lambda and K0s decays (called feed-down
correction).  The inverse corrections applied to the spectra for one
angular bin in the $tof$+$dE/dx$ analysis are shown in
Fig.~\ref{fig:tof-dedx-cor} as an example. The Systematic error
associated with each correction and with the particle identification
are also shown. The dominant systematic come from the uncertainty in
the correction for weak decays and secondary interactions
(30\% of the correction value). In addition to several track quality cuts,
maximum acceptance regions were selected by applying a cut on the
azimuthal angle, thereby assuring tracks have a large number of
measured points, and a very high reconstruction efficiency. This
minimizes the systematical errors arising from possible differences in
geometry between data and Monte-Carlo.

\begin{figure}[!h]
	\centering
        \includegraphics[width=.9\textwidth,height=.71\textheight]{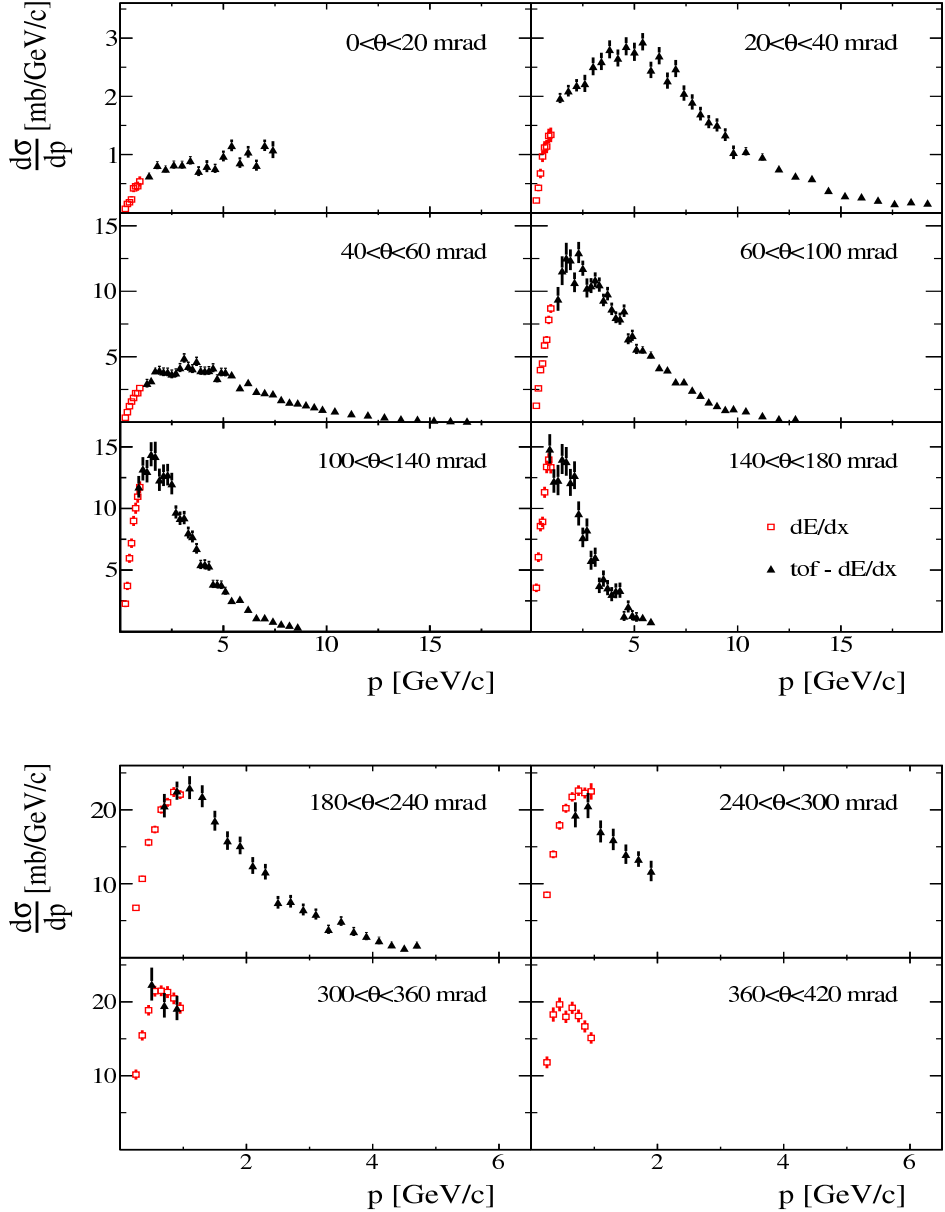}
	\caption{\label{pion_minus_all_mbarn}Differential cross sections for $\pi^{+}$ meson production
  in p+C interactions at 31~GeV/c. The spectra are presented as a
  function of laboratory  momentum ($p$)
  in different intervals of polar angle ($\theta$).
  Results obtained using two analysis methods are
  presented by different symbols:
  red open squares - $dE/dx$ analysis and black full triangles - $tof-dE/dx$ analysis.
  Error bars indicate only statistical uncertainties.}
\end{figure}

The spectra normalized to the inclusive cross section
\cite{claudia_pres} are shown in Fig.~\ref{pion_minus_all_mbarn} for
positively charged pions. The spectra are presented as a function of
particle momentum in ten intervals of the polar angle. The chosen
binning takes into account the available statistics of the 2007 data
sample, detector acceptance and particle production kinematics. The
negatively charged pion cross sections are given in \cite{pion_paper} along
with details on all three analysis.

\section{Conclusion}
The presented results are essential for precise predictions of
the neutrino flux in T2K and are currently used as input to the
neutrino beam simulation. In 2009 and 2010 another much larger set of
data has been collected with both the thin and a T2K replica carbon
target and is presently being analysed. For both these data sets the
ToF-F was extented yielding a higher detector acceptance, the TPC
readouts were upgraded and a new trigger system was implemented.
This new data will provide results of charged pion cross-section with a higher
precision and will allow the measurements of other hadron species such
as charged kaons, protons or $K^0_s$. Knowledge of kaon production is crucial
for T2K to predict the intrinsic $\nu_e$ contamination of the neutrino
beam.

\end{document}